\documentclass[12pt,draftclsnofoot,onecolumn]{IEEEtran} 
\ifCLASSINFOpdf
\else
\fi

\usepackage{graphicx,color}				
\usepackage{amssymb,amsmath,amsfonts,amsthm}  

\usepackage{algorithm,caption}
\usepackage[noend]{algpseudocode}

\def \tr {\text{tr}}

\begin{document}
%
\title{Integrated Semi-definite Relaxation Receiver for LDPC-Coded MIMO Systems}

\author{Kun~Wang,~\IEEEmembership{Member,~IEEE,}
        and~Zhi~Ding,~\IEEEmembership{Fellow,~IEEE}
\thanks{This work was partly presented in \cite{wang2018joint} (accepted for VTC 2018-Fall).}
}


%


\maketitle

\begin{abstract}
Semi-definite relaxation (SDR) detector has been demonstrated to be successful in 
approaching maximum likelihood (ML) performance while the time complexity is only polynomial. 
We propose a new receiver jointly utilizing the forward error correction (FEC) code information
 in the SDR detection process.
Strengthened by code constraints, the joint SDR detector substantially improves
 the overall receiver performance.
For further performance boost, the ML-SDR detector is adapted to MAP-SDR detector
by incorporating \textit{a priori} information in the cost function.
Under turbo principle, MAP-SDR detector takes in soft information from decoder 
and outputs extrinsic information with much improved reliability. 
We also propose a simplified SDR turbo receiver that solves only one SDR per codeword 
instead of solving multiple SDRs in the iterative turbo processing. 
This scheme significantly reduces the time complexity of SDR turbo receiver,
while the error performance remains similar as before. 
In fact, our simplified scheme is generic and it can be applied to any list-based iterative receivers.
\end{abstract}


%
\IEEEpeerreviewmaketitle

\section{Introduction}  \label{sec:intro}

Multiple-input multiple-output (MIMO) transceiver
technology represents a breakthrough in the advances of 
wireless communication systems. 
Modern wireless systems widely adopt multiple antennas, for example, the 3GPP LTE and WLAN systems \cite{molisch2012wireless},
and further massive MIMO has been proposed for next-generation wireless systems \cite{larsson2014massive}.
MIMO systems can provide manifold throughput increase, or can offer reliable transmissions by spatial diversity \cite{goldsmith2005wireless}. 
In order to fully exploit the advantages promised by MIMO, the 
receiver must be able to effectively recover the transmitted information.
Thus, detection and decoding remain to be one of the fundamental areas in state-of-the-art MIMO research.

It is well known that maximum likelihood (ML) detection is optimal in terms 
of minimum error probabilities for equally likely data sequence transmissions. 
However, the ML detection is NP-hard \cite{verdu1989computational} and its time complexity is exponential for MIMO detection, regardless of whether exhaustive search or other search algorithms (e.g., sphere decoding) 
are used \cite{jalden2005complexity} in data symbol detection. 
Aiming to reduce the high computational complexity for MIMO receivers,
a number of research efforts have focused on designing
near-optimal and high performance receivers. In the literature, 
the simplist linear receivers, such as matched filtering (MF), zero-forcing (ZF) and minimum mean squared error (MMSE), have been widely investigated. 
Other more reliable and more sophisticated receivers, such as
successive interference cancellation (SIC) or
parallel interference cancellation (PIC) receivers have also been
studied.  However, these receivers suffer substantial performance loss. 

In recent years, various semi-definite relaxation techniques have
emerged as a sub-optimum detection method that 
can achieve near-ML detection 
performance \cite{luo2010semidefinite}.
Specifically, ML detection of MIMO transmission 
can be formulated as least squares integer programming problem 
which can then be converted into an 
equivalent quadratic constrained quadratic program (QCQP). 
The QCQP can be transformed  by 
relaxing the rank-1 constraint into a semi-definite program.  
With the name \textit{semi-definite relaxation} (SDR), 
its substantial performance improvement over algorithms such as
MMSE and SIC has stimulated broad research interests as seen in
the works of
\cite{tan2001application,ma2002quasi,ma2004semidefinite,ma2006blind}. 
Several earlier works \cite{tan2001application,ma2002quasi} developed 
SDR detection in proposing multiuser detection for CDMA transmissions. 
Among them, the authors of \cite{ma2004semidefinite} proposed an SDR-based 
multiuser detector for $M$-ary PSK signaling. Another work in
\cite{ma2006blind} presented an efficient SDR 
implementation of blind ML detection 
of signals that utilize orthogonal space-time block codes. 
Furthermore, multiple SDR detectors of 16-QAM signaling 
were compared and shown to be equivalent in \cite{ma2009equivalence}. 

Although most of the aforementioned studies focused on SDR 
detections of uncoded transmissions, forward error correction (FEC)
codes in binary field have long been integrated into data communications
to effectively combat noises and co-channel interferences. 
Because FEC decoding takes place in the finite binary field whereas  
modulated symbol 
detection is formulated in the Euclidean space of complex field, 
the joint detection and decoding typically relies on the concept of
turbo processing. 
The authors of \cite{steingrimsson2003soft} 
simplified turbo receiver by reducing the number of 
optimization problems without performance loss. 
The authors of \cite{nekuii2011efficient} further developed
soft-output SDR receivers that are significantly less complex 
while suffering
only slight degradation than turbo receivers in performance.
More recently, as a follow-up paper of \cite{nekuii2011efficient}, the 
authors of \cite{salmani2017semidefinite} extended the 
efficient SDR receivers from 4-QAM (QPSK) to higher-order QAM signaling
by presenting two customized algorithms
for solving the SDR demodulators.

In this work, we present a new SDR detection algorithm for 
FEC coded MIMO systems with improved performance. 
In our design, FEC codes not only are used for decoding, but also are
integrated as constraints 
within the detection optimization formulation to 
develop a novel joint SDR detector
\cite{wang2014joint,wang2016robust,wang2015diversity}. 
Instead of using the more traditional randomization or rank-one 
approximation for symbol detection, our data detection takes
advantage of the last column of the optimal SDR matrix solution. 
We further propose a soft-in and soft-out SDR detector 
that demonstrates substantial performance
gain through iterative turbo processing. 
The proposed soft receiver has significantly lower 
complexity compared with the original full-list detector, 
while achieving similar bit error rate (BER) in overall performance. 
Furthermore,  we also present a simplified joint SDR turbo receiver. 
In this new approach, only one SDR is solved in the initial iteration
for each codeword, unlike existing works that require multiple SDR solutions. 
In subsequent iterations, we propose a simple approximation to
generate the requisite output extrinsic information for turbo message passing. 
In fact, our proposed approximation scheme is generic in the sense
that it can be jointly applied with other list-based iterative MIMO receivers. 
Compared with other SDR turbo receivers, 
both receivers in \cite{steingrimsson2003soft} and our new
work retain the original turbo detection performance.
More importantly, the complexity of our proposed scheme is much lower. 
On the other hand, the receivers presented 
in \cite{nekuii2011efficient} trades BER performance for 
low complexity. 


This manuscript is organized as follows. 
First, Section \ref{sec:sys_model} describes the baseband MIMO system model 
and a corresponding real field detection problem. 
We also present the SDR formulation to approximate the maximum likelihood
detection of MIMO signals.
In Section \ref{sec:code_anchor}, we integrate
the FEC code constraints into the SDR to form a joint ML-SDR receiver.
Section \ref{sec:turbo_sdr} incorporates the cost function of joint ML-SDR  
with \textit{a priori} information to develop a 
joint MAP-SDR problem to be tackled through turbo processing.  
We demonstrate the superior performance of joint SDR receivers in Section \ref{sec:sim}.
Finally, Section \ref{sec:con} concludes the work.

\section{System Model and SDR Detection}  \label{sec:sys_model}

\subsection{Maximum-likelihood MIMO Signal Detection}
Consider an $N_t$-input $N_r$-output spatial multiplexing MIMO system with memoryless channel.
The baseband equivalent model of this system at time $k$ can be expressed as
\begin{equation} \label{eq:mimo_complex}
\mathbf{y}_k^c = \mathbf{H}_k^c \mathbf{s}_k^c + \mathbf{n}_k^c, \quad k = 1, \ldots, K,
\end{equation}
where $\mathbf{y}_k^c \in \mathbb{C}^{N_r \times 1}$ is the received signal,
$\mathbf{H}_k^c \in \mathbb{C}^{N_r \times N_t}$ denotes the MIMO channel matrix,
$\mathbf{s}_k^c \in \mathbb{C}^{N_t \times 1}$ is the transmitted signal, and 
$\mathbf{n}_k^c  \in \mathbb{C}^{N_r \times 1}$ is an additive Gaussian noise vector, 
each element of which is independent and follows $\mathcal{CN}(0, 2 \sigma_n^2)$. 
In fact, besides modeling the point-to-point MIMO system, Eq.~(\ref{eq:mimo_complex}) can be also
used to model frequency-selective systems 
\cite{lathi2009modern}, multi-user systems \cite{wang2016fec}, among others. 
The only difference lies in the structure of channel matrix  $\mathbf{H}_k^c$.

To simplify problem formulation, the complex-valued
signal model can be transformed into
the real field by letting
\begin{equation*}
\mathbf{y}_k = 
\begin{bmatrix}
\text{Re}\{ \mathbf{y}_k^c  \} \\
\text{Im}\{ \mathbf{y}_k^c  \}
\end{bmatrix},
\mathbf{s}_k = 
\begin{bmatrix}
\text{Re}\{ \mathbf{s}_k^c  \} \\
\text{Im}\{ \mathbf{s}_k^c  \}
\end{bmatrix},
\mathbf{n}_k = 
\begin{bmatrix}
\text{Re}\{ \mathbf{n}_k^c  \} \\
\text{Im}\{ \mathbf{n}_k^c  \}
\end{bmatrix},
\end{equation*}
and
\begin{equation*}
\mathbf{H}_k = 
\begin{bmatrix}
\text{Re}\{ \mathbf{H}_k^c  \}  & - \text{Im}\{ \mathbf{H}_k^c  \}\\
\text{Im}\{ \mathbf{H}_k^c  \} & \text{Re}\{ \mathbf{H}_k^c  \}
\end{bmatrix}.
\end{equation*}
Consequently, the transmission equation is given by
\begin{equation} \label{eq:mimo_real}
\mathbf{y}_k = \mathbf{H}_k \mathbf{s}_k + \mathbf{n}_k, \quad k = 1, \ldots, K.
\end{equation}
In this study, we choose capacity-approaching LDPC code for the purpose of forward error correction. 
Further, we assume the transmitted symbols are generated based on QPSK constellation,
i.e., $s_{k,i}^c \in \{ \pm1 \pm j \}$ for $k = 1, \ldots, K$ and $i = 1, \ldots, N_t$.
The codeword (on symbol level) is placed first along the spatial dimension
and then along the temporal dimension.

Before presenting the code anchored detector, we begin with a brief review of existing 
SDR detector in uncoded MIMO systems for the convenience of subsequent integration.
By the above assumption of Gaussian noise, it can be easily shown that the optimal ML detection
is equivalent to the following discrete least squares problem
\begin{equation} \label{eq:ml_detection}
\underset{\mathbf{x}_k \in \{ \pm 1 \}^{2N_t}}{\text{min.}} \;
\sum_{k=1}^K \Vert \mathbf{y}_k - \mathbf{H}_k \mathbf{x}_k \Vert^2.
\end{equation}
However, this problem is NP-hard. Brute-force solution would take exponential time (exponential in $N_t$).
Sphere decoding was proposed for efficient computation of ML problem. 
Nonetheless, it is still exponentially complex, even on average sense \cite{jalden2005complexity}.

\subsection{SDR MIMO Detector}  \label{sec:uncoded_sdr}
SDR can generate an \textit{approximate} solution to the ML problem
in polynomial time. More specifically, the time complexity is $\mathcal{O}(N_t^{4.5})$ when a generic interior-point algorithm is used,
and it can be as low as $\mathcal{O}(N_t^{3.5})$ with a customized algorithm \cite{luo2010semidefinite}.
The trick of using SDR is to firstly turn the ML detection into a homogeneous QCQP 
by introducing auxiliary variables $\{t_k, k = 1, \ldots, K\}$ \cite{luo2010semidefinite}.
The ML problem can then be equivalently written as the following QCQP
\begin{equation} \label{eq:qcqp}
\begin{aligned}
& \underset{\{\mathbf{x}_k, t_k\}}{\text{min.}}
& &  \sum_{k=1}^K 
\begin{bmatrix}
\mathbf{x}_k^T & t_k
\end{bmatrix}
\begin{bmatrix}
\mathbf{H}_k^T \mathbf{H}_k & \mathbf{H}_k^T \mathbf{y}_k  \\
-\mathbf{y}_k^T \mathbf{H}_k & || \mathbf{y}_k ||^2 
\end{bmatrix} 
\begin{bmatrix}
\mathbf{x}_k \\
t_k
\end{bmatrix} \\
& \text{s.t.}
& & t_k^2 = 1, \; x_{k,i}^2 = 1, \; k = 1, \ldots, K, i = 1, \ldots, 2N_t.
\end{aligned}
\end{equation}

This QCQP is non-convex because of its quadratic equality constraints. 
To solve it approximately via SDR, define the rank-1 semi-definite matrix
\begin{equation} \label{eq:rank1_matrix}
\mathbf{X}_k = 
\begin{bmatrix}
\mathbf{x}_k \\
t_k
\end{bmatrix}
\begin{bmatrix}
\mathbf{x}_k^T & t_k
\end{bmatrix}
=
\begin{bmatrix}
\mathbf{x}_k \mathbf{x}_k^T & t_k \mathbf{x}_k \\
t_k \mathbf{x}_k^T & t_k^2
\end{bmatrix},
\end{equation}
and for notational convenience, denote the cost matrix by
\begin{equation} \label{eq:cost_matrix}
\mathbf{C}_k = 
\begin{bmatrix}
\mathbf{H}_k^T \mathbf{H}_k & \mathbf{H}_k^T \mathbf{y}_k  \\
-\mathbf{y}_k^T \mathbf{H}_k & || \mathbf{y}_k ||^2
\end{bmatrix}.
\end{equation}
Using the property of trace $\mathbf{v}^T\mathbf{Q}\mathbf{v} = \tr(\mathbf{v}^T\mathbf{Q}\mathbf{v}) = \tr(\mathbf{Q}\mathbf{v}\mathbf{v}^T)$,
the QCQP in Eq.~(\ref{eq:qcqp}) can be relaxed to SDR by removing the rank-1 constraint on $\mathbf{X}_k$.
Therefore, the SDR formulation is
\begin{equation} \label{eq:disjoint_sdr}
\begin{aligned}
& \underset{\{\mathbf{X}_k\}}{\text{min.}}
& &  \sum_{k=1}^K \tr(\mathbf{C}_k \mathbf{X}_k) \\
& \text{s.t.}
& & \tr(\mathbf{A}_i \mathbf{X}_k) = 1, \; k = 1, \ldots, K, i = 1, \ldots, 2N_t + 1, \\
& 
& & \mathbf{X}_k \succeq 0, \; k = 1, \ldots, K,
\end{aligned}
\end{equation}
where $\mathbf{A}_i$ is a zero matrix except that the $i$-th position on the diagonal is 1,
so $\mathbf{A}_i$ is used for extracting the $i$-th element on the diagonal of $\mathbf{X}_k$.
It is noted that $\mathbf{A}_i  \equiv \mathbf{A}_{i,k}, \forall k$; thus, the index $k$ is omitted
for $\mathbf{A}_{i,k}$ in Eq.~(\ref{eq:disjoint_sdr}). 
Finally, we would like to point out that the SDR problems formulated in most papers 
are targeted at a single time snapshot, since their system of interest is uncoded.
Here, for subsequent integration of code information,
we consider a total of $K$ snapshots that can accommodate an FEC codeword.


\section{FEC Codes in Joint SDR Receiver Formulation}  \label{sec:code_anchor}
If MIMO detector can provide more accurate 
information to downstream decoder, 
an improved decoding performance can be expected.
With this goal in mind, we propose to use FEC code information when performing detection.

\subsection{FEC Code Anchoring}
Consider an $(N_c,K_c)$ LDPC code. Let $\mathcal{M}$ and $\mathcal{N}$ be the 
index set of check nodes and variable nodes of the parity check matrix, respectively, i.e.,
$\mathcal{M} = \{1, \ldots, N_c-K_c \}$ and $\mathcal{N} = \{1, \ldots, N_c \}$.
Denote the neighbor set of the $m$-th check node as $\mathcal{N}_m$
and let $\mathcal{S} \triangleq \{ \mathcal{F} \, | \, \mathcal{F} \subseteq \mathcal{N}_m \, \text{with} \, |\mathcal{F}| \, \text{odd} \}$.
Then one characterization of fundamental polytope is captured by 
the following forbidden set (FS) constraints \cite{feldman2005using}
\begin{equation} \label{eq:parity_ineq}
\sum_{ n \in \mathcal{F} } f_n - \sum_{ n \in \mathcal{N}_m \backslash \mathcal{F}} f_n \leq |\mathcal{F}| - 1, \; \forall m \in \mathcal{M},
\forall \mathcal{F} \in \mathcal{S}
\end{equation}
plus the box constraints for bit variables
\begin{equation} \label{eq:box_ineq}
 0  \leq f_n \leq 1, \quad \forall n \in \mathcal{N}.
\end{equation}

Recall that the bits $\{f_n\}$ are mapped by modulators into
transmitted data symbols in $\mathbf{x}_k$. 
It is important to note that the parity check 
inequalities (\ref{eq:parity_ineq}) can help to tighten our 
detection solution of $\mathbf{x}_k$
by explicitly forbidding the bad configurations 
of $\mathbf{x}_k$ that are inconsistent with FEC codewords. 
Thus, a joint detection and decoding algorithm can take advantage
of these linear constraints by integrating them within the
SDR problem formualtion. 

Notice that coded bits $\{f_n\}$ are in fact binary. Hence, 
the box constraint of (\ref{eq:box_ineq}) is a relaxation of 
the binary constraints. 
In fact, if variables $f_n$'s are forced to be only 0's and 1's
(binary),  then the constraints (\ref{eq:parity_ineq}) will be equivalent 
to the original binary parity-check constraints.
To see this, if parity check node $m$ fails to hold, 
there must be a subset of variable nodes
$\mathcal{F} \subseteq \mathcal{N}_m$ of odd cardinality 
such that all nodes in $\mathcal{F}$
have the value 1 and all those in 
$\mathcal{N}_m \backslash \mathcal{F}$ have value 0.
Clearly, the corresponding parity inequality in (\ref{eq:parity_ineq}) 
would forbid such outcome.

\subsection{Symbol-to-Bit Mapping}
To anchor the FS constraints into the SDR formulation in Eq.~(\ref{eq:disjoint_sdr}),
we need to connect the bit variables $f_n$'s with the data vectors
$\mathbf{x}_k$'s or the matrix variables $\mathbf{X}_k$'s.

As stated in \cite{luo2010semidefinite}, if $(\mathbf{x}_k^*, t_k^*)$ is an optimal solution to (\ref{eq:disjoint_sdr}),
then the final solution should be $t_k^* \mathbf{x}_k^*$, where $t_k^*$ controls the sign of the symbol. 
In fact, Eq.~(\ref{eq:rank1_matrix}) shows that
the first $2 N_t$ elements of last column or last row are exactly $t_k \mathbf{x}_k$.
We also note that the first $N_t$ elements correspond to the real parts of the transmitted symbols
and the next $N_t$ elements correspond to the imaginary parts. 
Hence, for QPSK modulation,
the mapping constraints
for time instant $k = 1, \ldots, K$ are simply as follows
\begin{equation} \label{eq:qpsk_gray}
\begin{split}
& \tr(\mathbf{B}_i \mathbf{X}_k) = 1 - 2 f_{2N_t(k-1)+2i-1}, \; i = 1, \ldots, N_t, \\
& \tr(\mathbf{B}_{i+N_t} \mathbf{X}_k) = 1 - 2 f_{2N_t(k-1)+2i}, \; i = 1, \ldots, N_t,
\end{split}
\end{equation}
where $\mathbf{B}_i$ is a selection matrix
designed to extract the $i$-th element on the last column of $\mathbf{X}_k$:
\begin{equation}
\mathbf{B}_i = 
\begin{bmatrix}
0 & \ldots & \ldots & \ldots & 0 \\
\vdots   & \ddots &  & & \vdots  \\
\vdots   & & 0 &  & 1 \\
\vdots  & & & \ddots & \vdots \\
0 & \ldots & 0 & \ldots & 0
\end{bmatrix}, \; 1 \leq i \leq 2 N_t.
\end{equation}
The non-zero entry of $\mathbf{B}_i$ is the $i$-th element on the last column. 
For the same reason as that of $\mathbf{A}_i$,
the index $k$ is omitted in $\mathbf{B}_i$.
Moreover, note the subtle difference 
that $\mathbf{A}_i$ is defined for $1 \leq i \leq 2N_t + 1$ 
while $\mathbf{B}_i$ is defined for $1 \leq i \leq 2N_t$.

\subsection{Joint ML-SDR Receiver}
Having defined the necessary notations and constraints, 
a joint ML-SDR detector can be formulated as the following optimization
problem:
\begin{equation} \label{eq:joint_ml_sdr}
\begin{aligned}
& \underset{\{\mathbf{X}_k, f_n\}}{\text{min.}}
& &  \sum_{k=1}^K \tr(\mathbf{C}_k \mathbf{X}_k) \\
& \text{s.t.}
& & \tr(\mathbf{A}_i \mathbf{X}_k) = 1, \, \mathbf{X}_k \succeq 0, \quad k = 1, \ldots, K, i = 1, \ldots, 2N_t+1, \\
&
& & \tr(\mathbf{B}_i \mathbf{X}_k) = 1 - 2 f_{2N_t(k-1)+2i-1}, \quad k = 1, \ldots, K, i = 1, \ldots, N_t, \\
&
& & \tr(\mathbf{B}_{i+N_t} \mathbf{X}_k) = 1 - 2 f_{2N_t(k-1)+2i}, \quad k = 1, \ldots, K, i = 1, \ldots, N_t, \\
&
& & \sum_{ n \in \mathcal{F} } f_n - \sum_{ n \in \mathcal{N}_m \backslash \mathcal{F}} f_n \leq |\mathcal{F}| - 1, \quad \forall m \in \mathcal{M}, \forall \mathcal{F} \in \mathcal{S};\\ 
& &&
0 \leq f_n \leq 1, \quad \forall n \in \mathcal{N}.
\end{aligned}
\end{equation}

Recall that the matrix $\mathbf{X}_k\succeq 0 $ is a relaxation of
the rank one matrix 
\[
\mathbf{X}_k = 
\begin{bmatrix}
	\mathbf{x}_k \\
	t_k
\end{bmatrix}
\begin{bmatrix}
	\mathbf{x}_k^T & t_k
\end{bmatrix}\]
After obtaining the optimal solution $\{ \mathbf{X}_k \}$ of the SDR, one must
determine the final detected symbol values in $\mathbf{x}_k$. 
Traditionally, one
``standard'' approach to retrieve the final solution
is via Gaussian randomization that views $\mathbf{X}_k$ as the covariance matrix
of $\mathbf{x}_k$. Another method is to apply
rank-one approximation of $\mathbf{X}_k$ \cite{luo2010semidefinite}.

However, a more convenient way is to directly use the first $2N_t$ elements in the last column of $\mathbf{X}_k$.
If hard-input hard-output decoding algorithm (such as bit flipping) is used, 
we can first quantize $t_k^* \mathbf{x}_k^*$ into binary values before feeding 
them to the FEC decoder for error correction. 
On the other hand, for soft-input soft-output decoder 
such as sum-product algorithm (SPA), log-likelihood ratio (LLR) can be generated 
from the unquantized $t_k^* \mathbf{x}_k^*$.
Here, we caution that the unquantized results from Gaussian randomization are not suitable for soft decoders such as the SPA, because the magnitudes of the
LLRs generated from randomization do not accurately reflect the data bits' 
actual reliability. 



\section{Iterative Turbo SDR}  \label{sec:turbo_sdr}
\begin{figure}[!tb]
\centering
\centerline{\includegraphics[width=12cm]{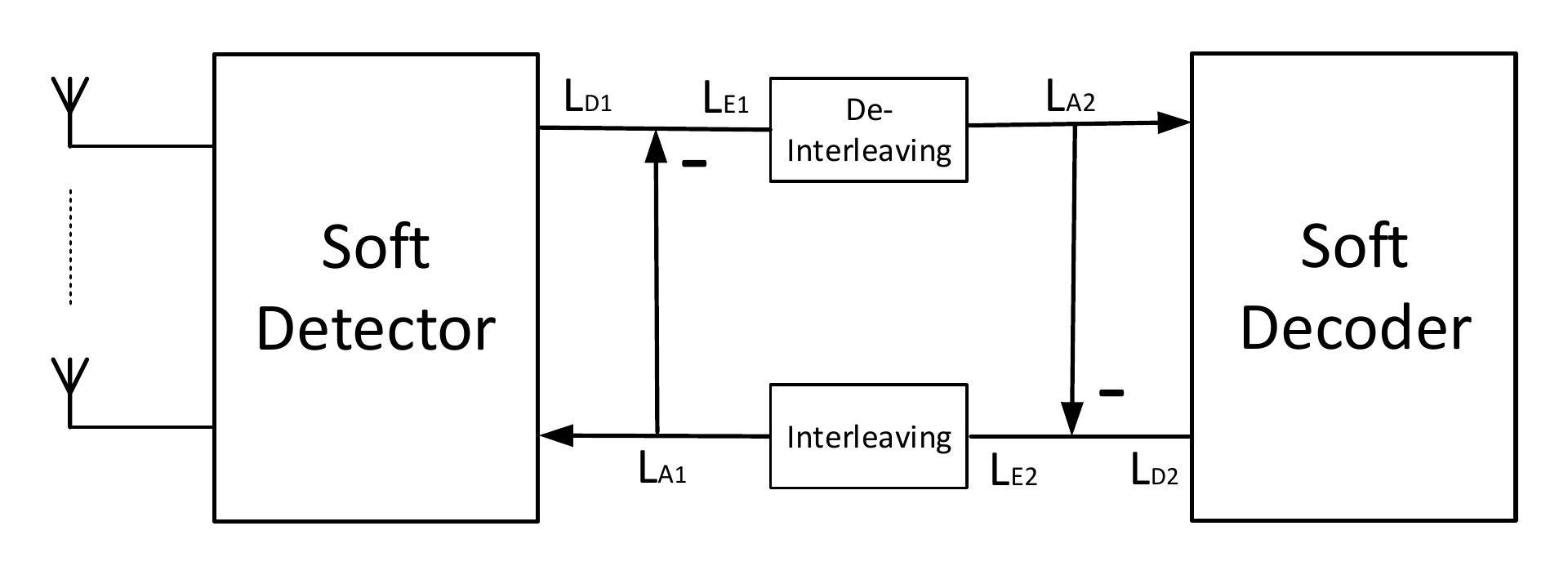}}
\caption{\small{Structure of Turbo Receiver.}}
\label{fig:turbo_receiver}
\vspace{-3mm}
\end{figure}

As demonstrated in \cite{hochwald2003achieving}, turbo receiver with iterative detection and decoding is capacity-approaching. 
Inspired by the turbo concept, we present an iterative SDR processing built upon the proposed joint ML-SDR
in Eq.~(\ref{eq:joint_ml_sdr}). 
The structure of turbo receiver is shown in Fig.~\ref{fig:turbo_receiver},
where our design focus is the soft detector that 
takes in priori LLRs from decoder and generates the posterior LLR of each interleaved bit for decoding, 
while the decoder uses standard SPA.

\subsection{Joint MAP-SDR Receiver}
When \textit{a priori} information of each bit is available, we can employ \textit{maximum a posterior} (MAP) criterion 
instead of ML. Specifically according to \cite{hagenauer1997turbo}, we have
\begin{align}
p(\mathbf{y}_k | \mathbf{s}_k) & \propto \exp(-|| \mathbf{y}_k - \mathbf{H}_k \mathbf{s}_k ||^2/(2 \sigma_n^2))\\
p\left(\mathbf{s}_k = \mathcal{M}(\mathbf{b}_k)\right)& \propto \exp(\mathbf{L}_{A1,k}^T \mathbf{b}_k / 2)
\end{align}
where $\mathbf{s}_k = \mathcal{M}(\mathbf{b}_k)$ denotes the modulator applied on a vector of polarized bits ($\pm1$),
and $\mathbf{L}_{A1,k}$ is the prior LLR vector corresponding to $\mathbf{b}_k$.
Here, we note that the polarized bit $b_{i,k} = 1 - 2 c_{i,k}$ for coded bit $c_{i,k} \in \{0,1\}$.
Therefore, the \textit{a posterior} probability is given as
\begin{equation}
p(\mathbf{s}_k | \mathbf{y}_k) 
\propto p(\mathbf{y}_k | \mathbf{s}_k) p(\mathbf{s}_k) 
\propto \exp(-|| \mathbf{y}_k - \mathbf{H}_k \mathbf{s}_k ||^2/(2 \sigma_n^2) + \mathbf{L}_{A1,k}^T \mathbf{b}_k / 2).
\end{equation}
After taking logarithm and summing over the $K$ time instants, MAP is equivalent to minimizing the new cost function
\begin{align} 
\sum_{k=1}^K \tr(\mathbf{C}_k \mathbf{X}_k) - \sigma_n^2 \mathbf{L}_{A1}^T (\mathbf{1} - 2 \mathbf{f})
 = \sum_{k=1}^K \tr(\mathbf{C}_k \mathbf{X}_k) + 2 \sigma_n^2 \mathbf{L}_{A1}^T \mathbf{f}.
 \end{align}
Therefore, the following optimization problem describes 
the new joint MAP-SDR detector 
\begin{equation} \label{eq:joint_map_sdr}
\begin{aligned}
& \underset{\{\mathbf{X}_k, f_n\}}{\text{min.}}
& &  \sum_{k=1}^K \tr(\mathbf{C}_k \mathbf{X}_k) + 2 \sigma_n^2 \mathbf{L}_{A1}^T \mathbf{f} \\
& \text{s.t.}
& & \tr(\mathbf{A}_i \mathbf{X}_k) = 1, \, \mathbf{X}_k \succeq 0, \quad k = 1, \ldots, K, i = 1, \ldots, 2N_t+1, \\
&
& & \tr(\mathbf{B}_i \mathbf{X}_k) = 1 - 2 f_{2N_t(k-1)+2i-1}, \quad k = 1, \ldots, K, i = 1, \ldots, N_t, \\
&
& & \tr(\mathbf{B}_{i+N_t} \mathbf{X}_k) = 1 - 2 f_{2N_t(k-1)+2i}, \quad k = 1, \ldots, K, i = 1, \ldots, N_t, \\
&
& & \sum_{ n \in \mathcal{F} } f_n - \sum_{ n \in \mathcal{N}_m \backslash \mathcal{F}} f_n \leq |\mathcal{F}| - 1, \quad \forall m \in \mathcal{M}, \forall \mathcal{F} \in \mathcal{S}; \\
&&& 0  \leq f_n \leq 1, \quad \forall n \in \mathcal{N}.
\end{aligned}
\end{equation}

\noindent{\bf Remark:}
Notice that our MAP cost function in Eq.~(\ref{eq:joint_map_sdr}) is 
generally applicable to any QAM constellations,
whereas the approach in \cite{salmani2017semidefinite} 
was to approximate the cost function for higher order QAM.
For higher order QAM beyond QPSK, the necessary changes 
for our joint SDR receiver include box relaxation of diagonal elements of  $\mathbf{X}_k$ \cite{ma2009equivalence}
and the symbol-to-bit mapping constraints.
We refer interested readers to our previous works \cite{wang2015joint,wang2016diversity,wang2016fec} for the details of 
higher order QAM mapping constraints.

Let the vector with superscript $[i]$ denote a vector excluding the $i$-th element.
Also, denote $\mathcal{L} = \{-1,+1\}^{2N_t}$ and $\mathcal{L}_{i,\pm1} = \left\{ \mathbf{b} \in \mathcal{L} \, | \, b_i = \pm 1 \right\}$. 
Following the derivations in \cite{hochwald2003achieving}, 
the extrinsic LLR of bit $b_{i,k}$ (the $i$-th bit at time $k$) with max-log approximation is given by
\begin{equation} \label{eq:extr_llr}
\begin{split}
L_{E1} (b_{i,k}) \approx
\max_{\mathbf{b}_k \in \mathcal{L}_{i,+1}} \left\{ -\frac{|| \mathbf{y}_k - \mathbf{H}_k \mathbf{s}_k ||^2}{2 \sigma_n^2} + 
\frac{(\mathbf{L}_{A1,k}^{[i]})^T \mathbf{b}_{k}^{[i]}}{2} \right\} \\
- \max_{\mathbf{b}_k \in \mathcal{L}_{i,-1}} \left\{ -\frac{|| \mathbf{y}_k - \mathbf{H}_k \mathbf{s}_k ||^2}{2 \sigma_n^2} + 
\frac{(\mathbf{L}_{A1,k}^{[i]})^T \mathbf{b}_{k}^{[i]}}{2} \right\} 
\end{split}
\end{equation}

It is noted that the cardinality of $\mathcal{L}$ is exponential in $N_t$.
With the solution from our joint MAP-SDR detector, it is unnecessary to enumerate over the full list $\mathcal{L}$.
Instead, we can construct a subset $\mathcal{\overline{L}}_k \subseteq \mathcal{L}$, containing the 
probable candidates that are within a certain Hamming distance from the SDR optimal solution $\mathbf{b}_k^*$ \cite{love2005space}.
More specifically, $\mathcal{\overline{L}}_k = 
\left\{ \mathbf{b}'_k \in \mathcal{L} \, | \, d(\mathbf{b}'_k, \mathbf{b}_k^*) 
\leq P  \right\}$,
where the Hamming distance $d(\mathbf{b}', \mathbf{b}'') = \text{card}( \{i \, | \, b'_i \neq b''_i \})$. 
Correspondingly, we have $\mathcal{\overline{L}}_{i,k,\pm1} = \left\{ \mathbf{b}_k \in \mathcal{\overline{L}}_k \, | \, b_{i,k} = \pm 1 \right\}$.
The radius $P$ determines the cardinality of 
$\mathcal{\overline{L}}_k$, 
that is, $\text{card}(\mathcal{\overline{L}}_k) = 
\sum_{j=0}^P \binom{2N_t}{j}$.
Compared to the full list's size $2^{2N_t}$, this could significantly reduce the list size with the selection of small $P$.

We now briefly summarize the steps of this novel turbo receiver: 
\begin{description}
	\item[S0] To initialize, let the first iteration  
$\mathbf{L}_{A1} = \mathbf{0}$, and select a value $P$. 
	\item[S1] Solve the joint MAP-SDR given in Eq.~(\ref{eq:joint_map_sdr}).
	\item[S2] Generate a list $\mathcal{\overline{L}}_k$ with a given $P$, and generate extrinsic LLRs $\mathbf{L}_{E1}$ via Eq.~(\ref{eq:extr_llr}) with $\mathcal{L}_{i,\pm1}$ being replaced by $\mathcal{\overline{L}}_{i,k,\pm1}$.
	\item[S3] Send $\mathbf{L}_{E1}$ to SPA decoder. If maximum 
iterations are reached or if all FEC parity checks are satisfied after decoding, 
stop the turbo process; Otherwise, return to S1.
\end{description}

\subsection{Simplified Turbo SDR Receiver}

One can clearly see that it is costly for our proposed turbo SDR
algorithm to solve one joint MAP-SDR in each iteration (in step S1).
To reduce receiver complexity, 
we can actually solve one joint MAP-SDR in the first iteration
(i.e., the joint ML-SDR) 
and generate the candidate list by other means in subsequent iterations
without repeatedly solving the joint MAP-SDR. 
In fact, the authors \cite{nekuii2011efficient} proposed 
a Bernoulli randomization method to generate such a candidate list 
based only on the first iteration SDR output and subsequent decoder feedback.
We now propose another list generation method for our receiver
that is more efficient.

\begin{figure*}[!tb]
	    \begin{minipage}[b]{0.5\linewidth}
	      \centering
	      \centerline{\includegraphics[width=6cm]{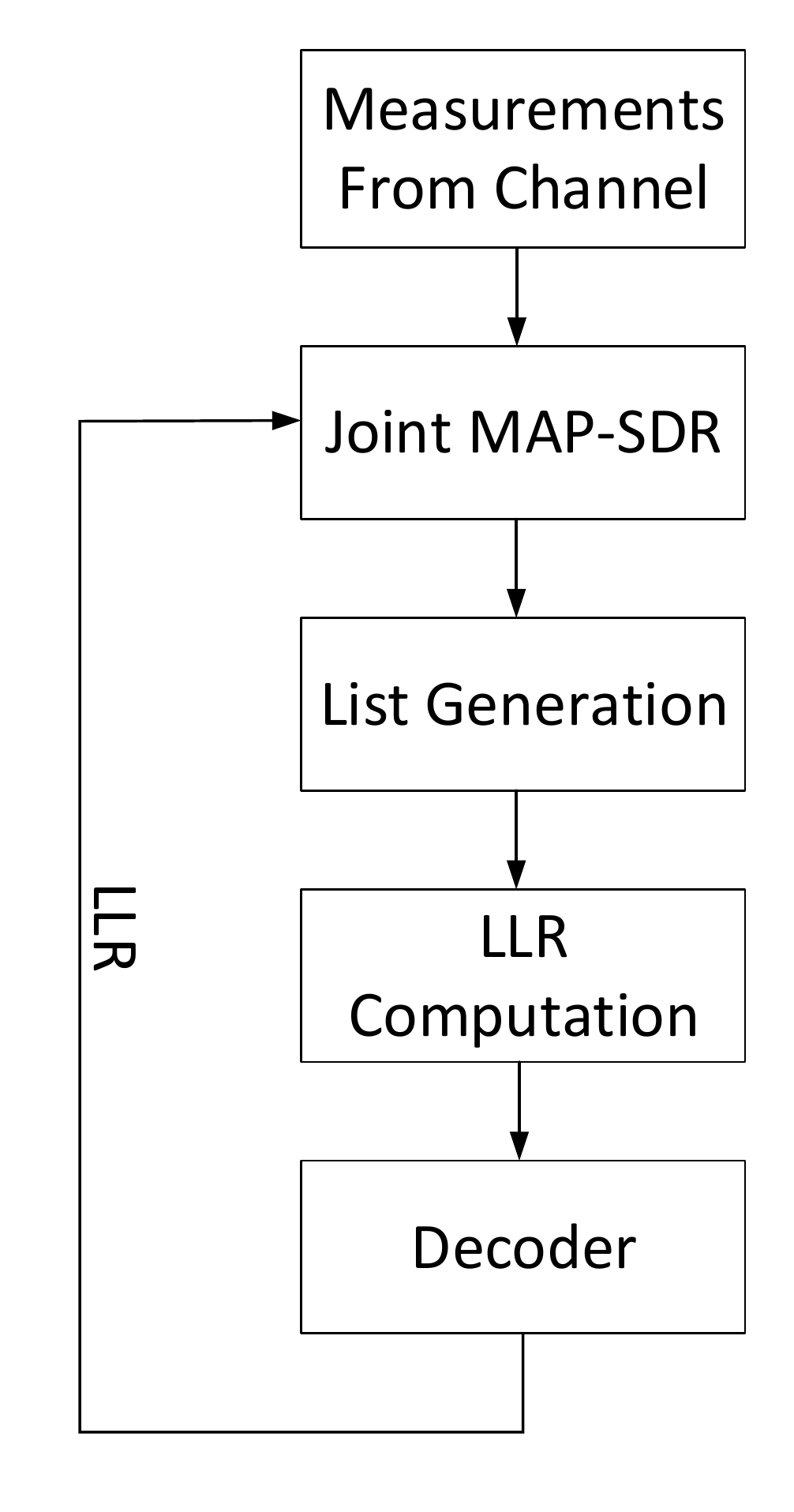}}
	      \centerline{(a)}\medskip
	    \end{minipage}
	    \hfill
	    \begin{minipage}[b]{.5\linewidth}
	      \centering
	      \centerline{\includegraphics[width=6cm]{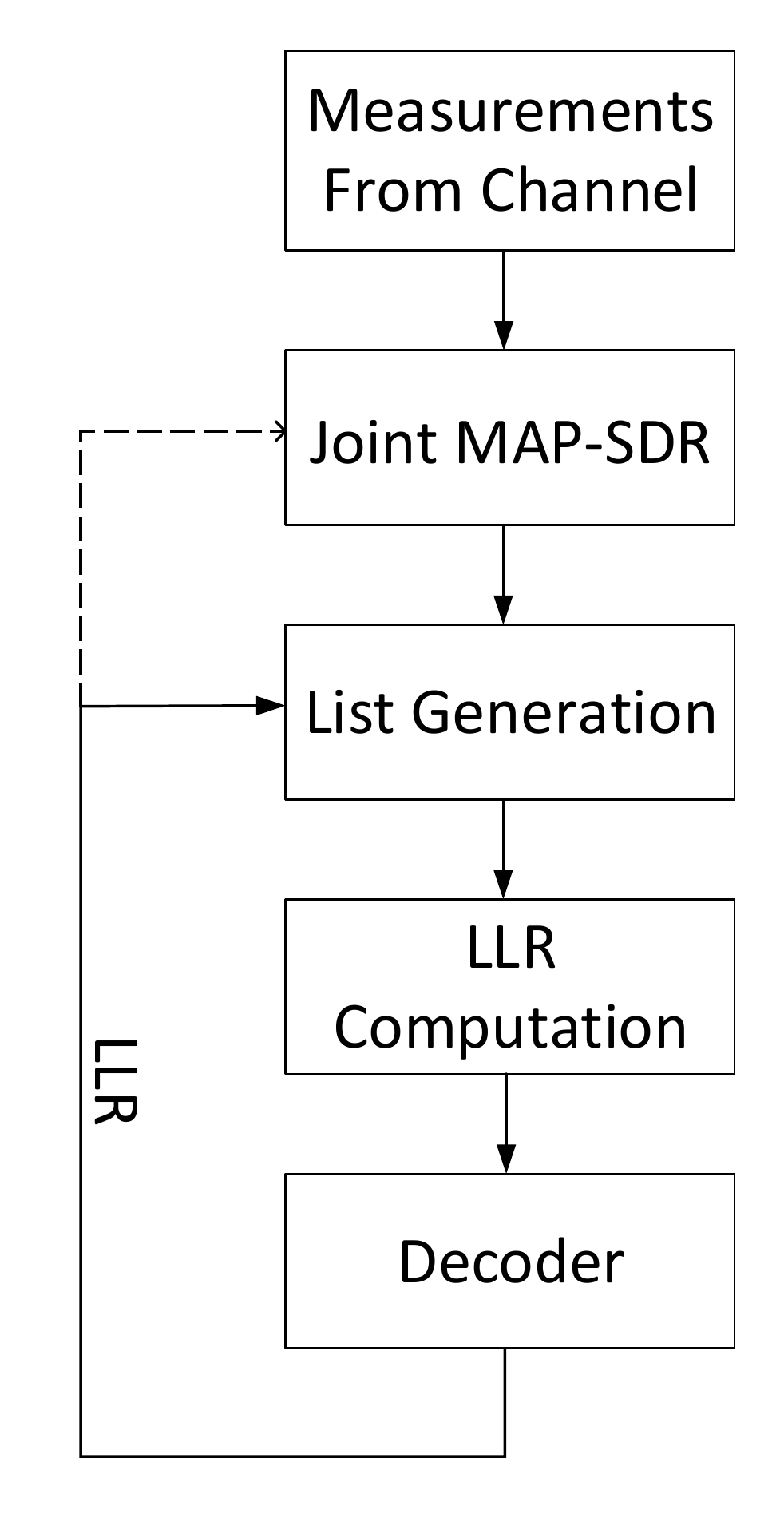}}
	      \centerline{(b)}\medskip
	    \end{minipage}
	    \caption{\small{(a) Flow of Multi Joint SDR. (b) Flow of Single Joint SDR.}}
	    \vspace*{-5mm}
	    \label{fig:flowcharts}
	\end{figure*}

The underlying principle of turbo receiver is that soft detector should use information from both received signals and decoder feedback
to improve receiver performance from one iteration to another. 
During the initial iteration, we solve the joint ML-SDR as shown in Eq.~(\ref{eq:joint_ml_sdr}).
The extrinsic LLR from this first iteration is denoted as $\mathbf{L}_{E1}^{init}$, 
which corresponds to the information that can be extracted from received signals.
When \textit{a priori} LLR value $\mathbf{L}_{A1}$ becomes available 
after the first iteration,
we combine them directly 
as $\mathbf{L}_{E1}^{comb} = \mathbf{L}_{E1}^{init} + \mathbf{L}_{A1}$,
and perform hard decision on $\mathbf{L}_{E1}^{comb}$ to obtain the bit vector $\mathbf{b}_k^*$ 
for each snapshot $k$. 
We then can generate list $\mathcal{\hat{L}}_k$ as before 
according to a pre-specified $P$.
The comparison with multiple-SDR turbo receiver is 
illustrated by flowcharts in Fig.~\ref{fig:flowcharts}.

We note that $\mathbf{L}_{A1}$ varies 
from iteration to iteration,
as does $\mathbf{L}_{E1}^{comb}$.
If $\mathbf{L}_{A1}$ converges towards a ``good solution'', it 
would enhance $\mathbf{L}_{E1}^{comb}$.
If $\mathbf{L}_{A1}$ is moving towards a ``poor solution'', 
then the initial LLR $\mathbf{L}_{E1}^{init}$ should help readjust
$\mathbf{L}_{E1}^{comb}$ to certain extent. 
In particular, if the joint ML-SDR detector 
(in the first iteration) can provide
a reliably good starting point $\mathbf{L}_{E1}^{init}$ for the turbo receiver,
then additional information that is extracted from 
resolving MAP-SDR in subsequent iterations can be quite limited. 
As will be shown in our simulations, this simple receiver scheme can
generate output performance that is close to the
original algorithm that requires solving joint MAP-SDR 
in each iteration.

\section{Simulation Results} \label{sec:sim}
In the simulation tests, a MIMO system with $N_t = 4$ and $N_r = 4$ is assumed.
The MIMO channel coefficients are assumed to be ergodic Rayleigh fading.
QPSK modulation is used and a regular (256,128) LDPC code with column weight 3 is employed; unless stated otherwise.

\subsection{ML-SDR Receiver Performance}
\begin{figure}[!htb]
\centering
\centerline{\includegraphics[width=12cm]{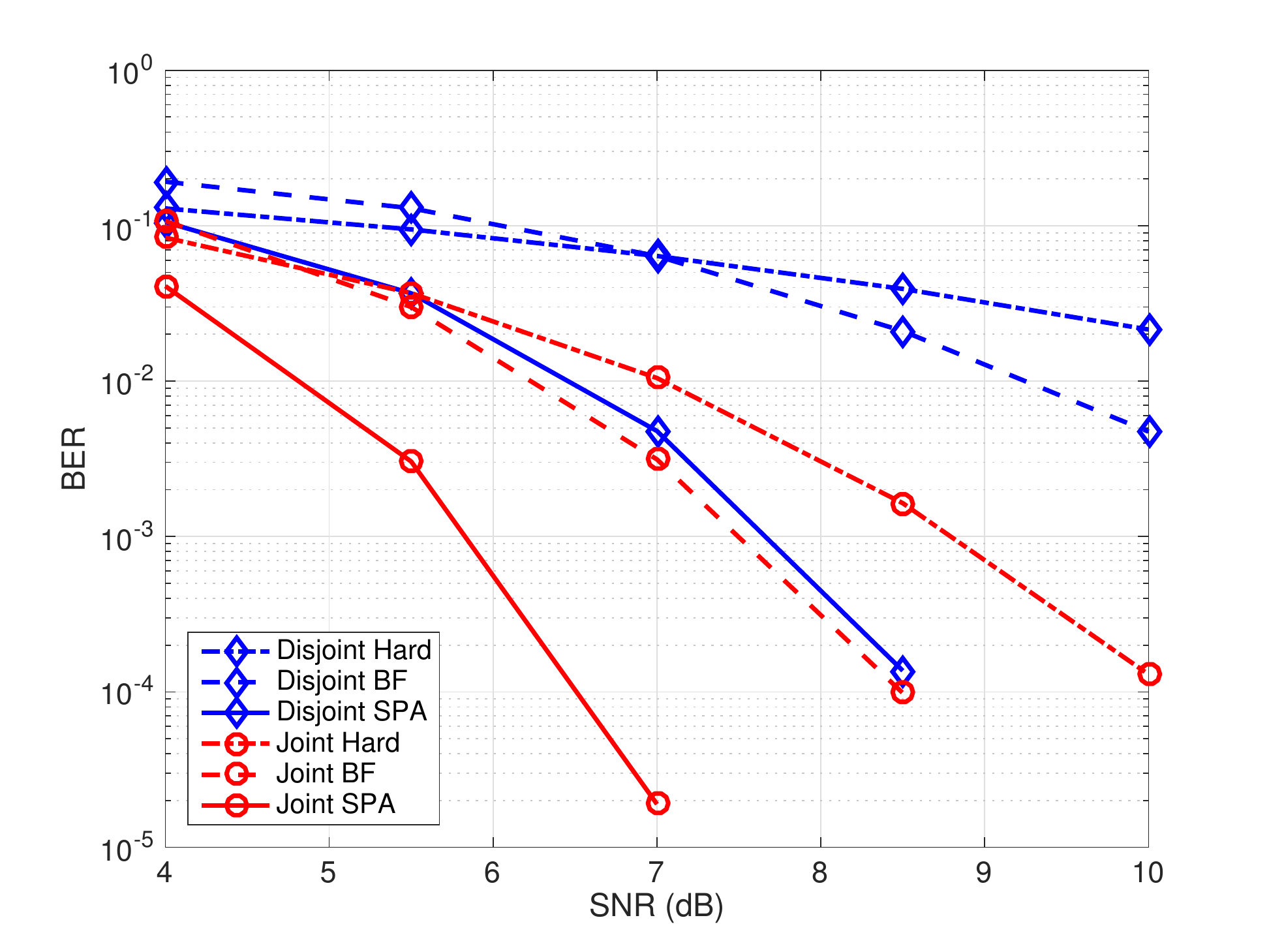}}
\caption{\small{BER comparisons of disjoint and joint SDR receivers: Rank 1 approximation.}}
\label{fig:rank1}
\end{figure}

\begin{figure}[!htb]
\centering
\centerline{\includegraphics[width=12cm]{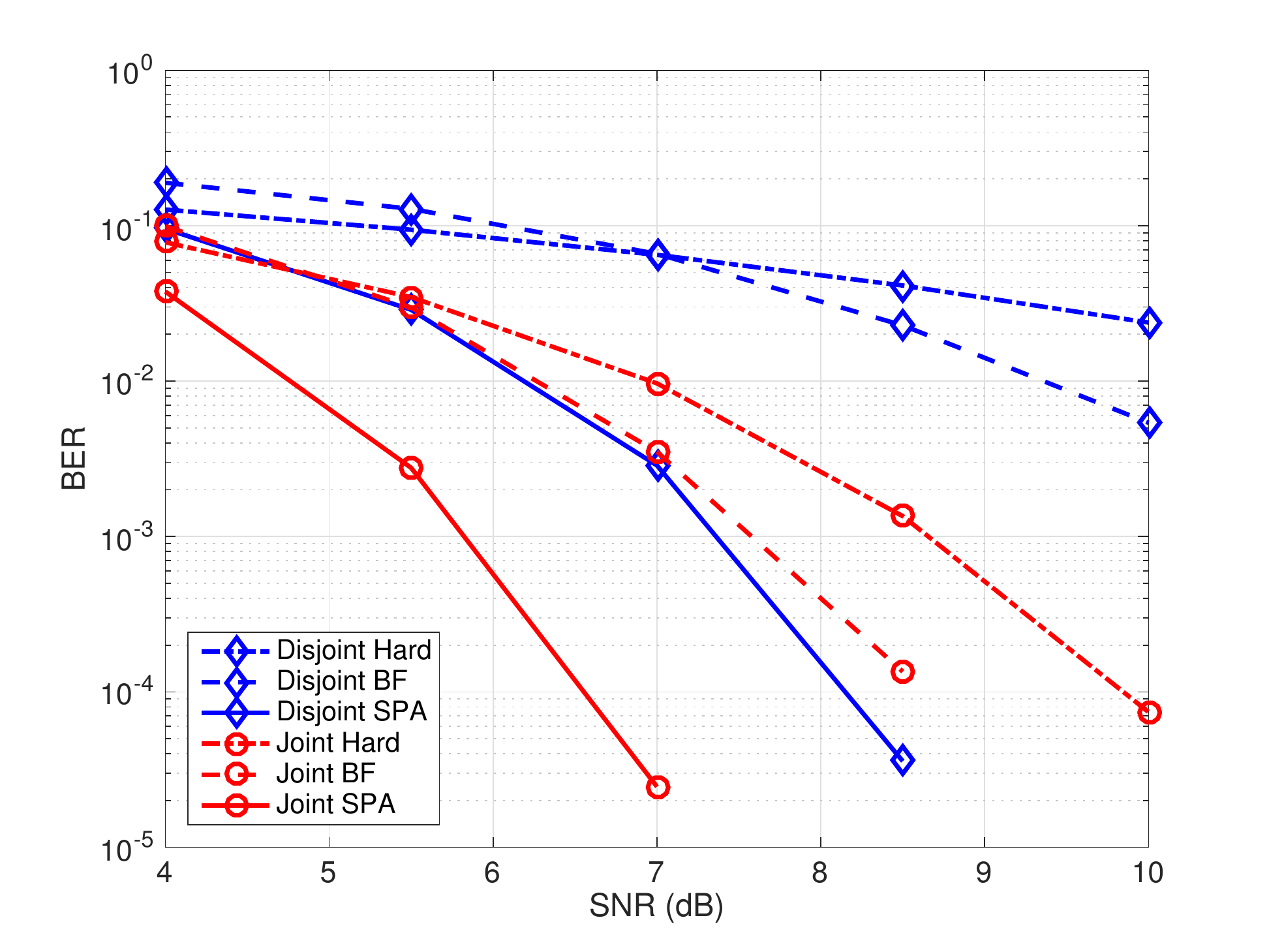}}
\caption{\small{BER comparisons of disjoint and joint SDR receivers: Direct
		approach using the final column of $\mathbf{X}$.}}
\label{fig:lastcol}
\end{figure}

In this subsection, we will demonstrate the power of code anchoring. 
We term the formulation in Eq.~(\ref{eq:disjoint_sdr}) as \textit{disjoint ML-SDR}, 
while that in Eq.~(\ref{eq:joint_ml_sdr}) as \textit{joint ML-SDR}.
With the optimal SDR solution $\{\mathbf{X}_k^*\}$, there are several approaches to retrieve 
the final solution $\mathbf{\hat{s}}_k$.
\begin{enumerate}
\item[-] \textit{Rank-1 approximation}: Perform eigen-decomposition on $\mathbf{X}_k^*$ to obtain the largest eigenvalue $e_k$ and its corresponding eigenvector $\mathbf{v}_k$. The final solution $\mathbf{\hat{s}}_k = \sqrt{e_k} \mathbf{v}_k[1:2N_t] \times \mathbf{v}_k[2N_t+1]$,
where we use slicing operation on vectors.
\item[-] \textit{Direct approach}: The final solution is retrieved from the last column of $\mathbf{X}_k$, 
i.e., $\mathbf{\hat{s}}_k = \mathbf{X}_k[1:2N_t, 2N_t+1]$.
\item[-] \textit{Randomization}: Generate $\mathbf{v}_k \sim \mathcal{CN}(\mathbf{0}, \mathbf{X}_k)$ for a certain number of trials, and pick the one that results in smallest cost value. Note that when evaluating the cost value, the elements of $\mathbf{v}_k$ are quantized to $\{-1,+1\}$.  
\end{enumerate}
We caution that, among the methods mentioned above, randomization is not suitable for soft decoding,
because the magnitudes of the randomized symbols do not reflect the actual reliability level.
Therefore, in the following, we will only consider rank-1 approximation and direct method,
the BER curves of which are shown in Fig.~\ref{fig:rank1} and Fig.~\ref{fig:lastcol}, respectively.
In the performance evaluation, we consider 1) hard decision on symbols, 2) bit flipping (BF) decoding
and 3) SPA decoding. 
In some sense, hard decision shows the ``pure'' gain by incorporating code constraints. 
BF is a hard decoding algorithm that performs moderately and SPA using LLR is the best.
If we compare the SPA curves within each figure, the SNR gain is around 2 dB at BER = 1e-4.
For other curves, the gains are even larger.
On the other hand, if we compare the curves across the two figures, their performances are quite similar.
Therefore, we do not need an extra eigen-decomposition; the direct approach is just as good.

\subsection{Joint MAP-SDR Turbo Receiver Performance}
\begin{figure}[!htb]
\centering
\centerline{\includegraphics[width=12cm]{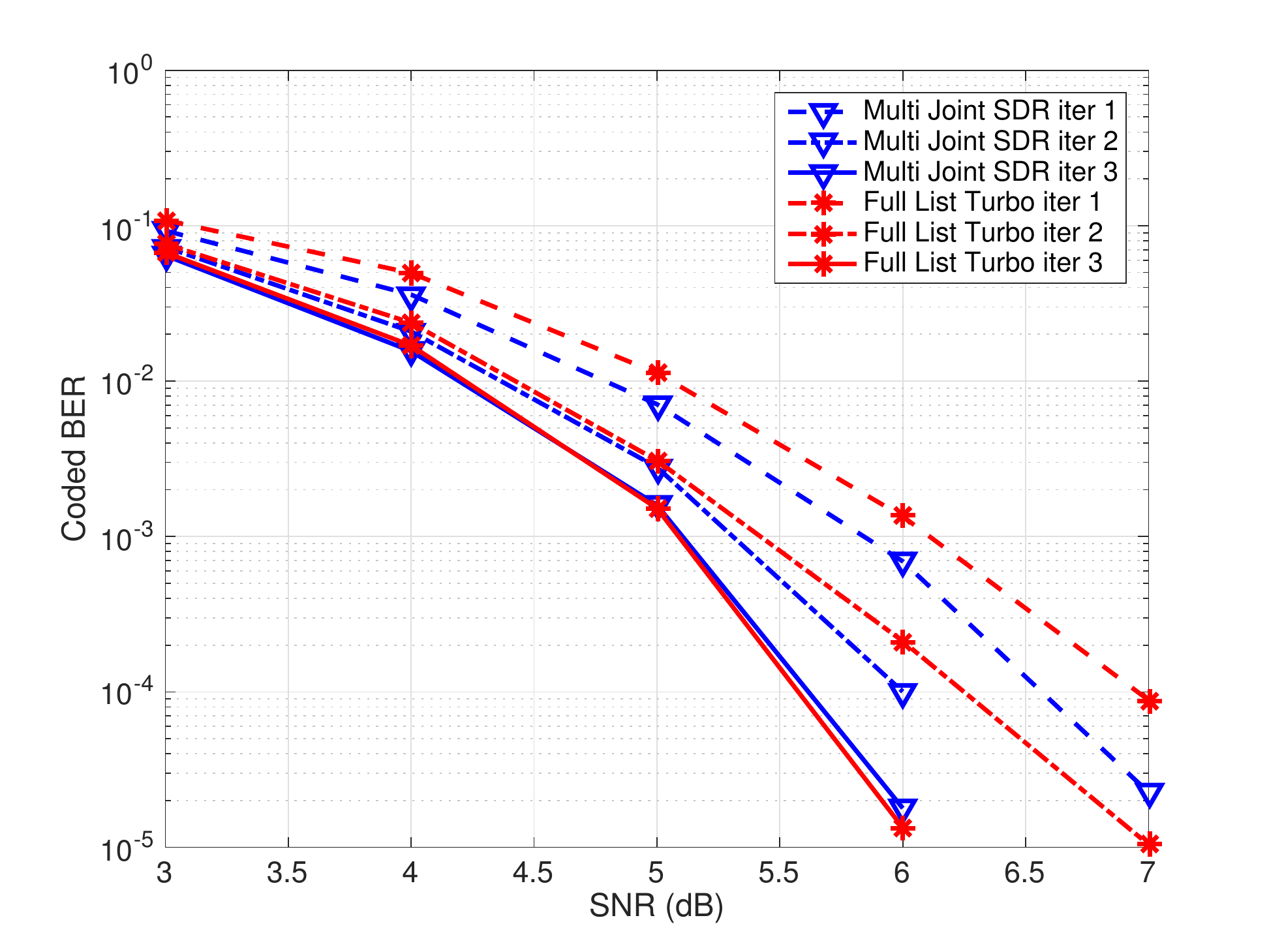}}
\caption{\small{BER comparisons of turbo equalizer and iterative SDR receiver at different iterations.}}
\label{fig:iter_comp}
\end{figure}

We investigate the performance of joint MAP-SDR turbo receiver versus full list turbo receiver.
In this test, we are more focused on the performance aspect with less concern on complexity, 
therefore we choose to run joint MAP-SDR in each iteration.
We name this turbo receiver \textit{multi joint SDR} in the figure legend.
We set Hamming radius $P=2$ and clipping value 8 for $\mathbf{L}_{E1}$.
Fig.~\ref{fig:iter_comp} shows the BER performance of 1st, 2nd and 3rd iterations. 
It is clear that joint MAP-SDR produces even better results than full list turbo in the 1st iteration. 
In later iterations, their performances gradually become the same.
We comment that the superior performance of MAP-SDR in the 1st iteration is
 because the maximizer in subset $\mathcal{\overline{L}}_k$ could be the ``true'' maximizer
 whereas the maximizer in set $\mathcal{L}$ might not be the true one due to noise perturbation. 

\begin{figure}[!htb]
\centering
\centerline{\includegraphics[width=12cm]{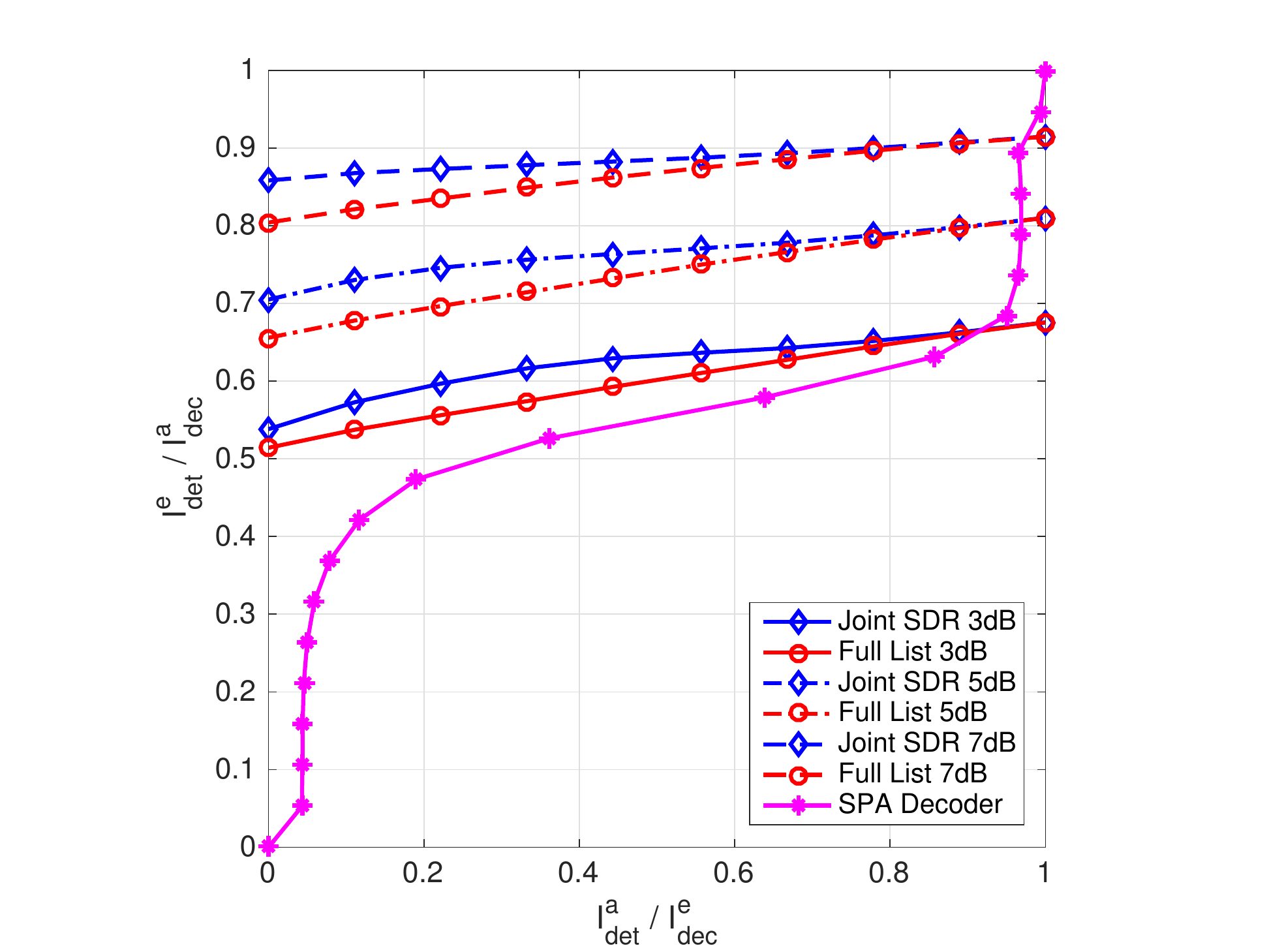}}
\caption{\small{EXIT charts of turbo equalizer and iterative SDR receiver at different SNRs.}}
\label{fig:exit_comp}
\end{figure}

We also plot the extrinsic information transfer (EXIT) charts of turbo receivers that are based on joint MAP-SDR 
and full list in Fig.~\ref{fig:exit_comp} to corroborate the BER performance at various SNRs. 
Here we use the histogram method to measure the extrinsic information \cite{rob}.
When \textit{a priori} mutual information (MI) is low, the output MI of joint MAP-SDR is much higher than that of full list. 
As iteration goes, MI becomes higher, and their gap becomes smaller.

\subsection{Simplified SDR Turbo Receiver Performance}
\begin{figure}[!htb]
\centering
\centerline{\includegraphics[width=12cm]{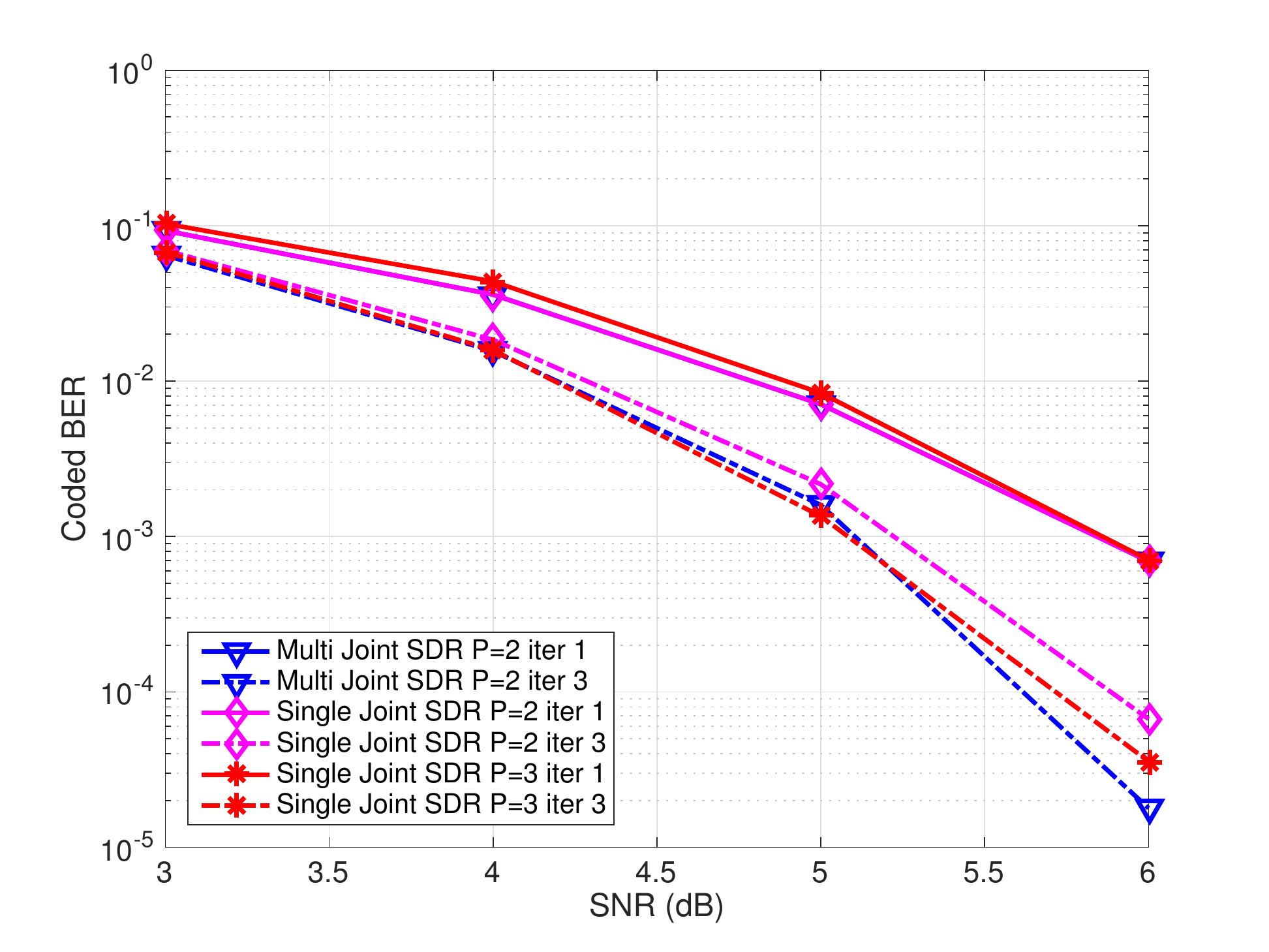}}
\caption{\small{BER comparisons of single SDR turbo receivers at different iterations.}}
\label{fig:single_sdr}
\end{figure}

The performance of \textit{single joint SDR} turbo receiver, which only runs joint MAP-SDR receiver in the initial iteration, 
is shown in Fig.~\ref{fig:single_sdr} in comparison with the \textit{multi joint SDR} that runs joint MAP-SDR in each iteration.
We choose two Hamming radii $P = 2$ and 3 for single joint SDR, while that for multi joint SDR is fixed at 2.
It is clear that they all perform equally good in the first iteration since the same joint MAP-SDR is invoked in that iteration.
At the 3rd iteration, single joint SDR slightly degrades, especially for $P=2$, but the performance degradation is acceptable 
in trade for such low complexity.

\subsection{Comparison with Other SDR Receivers}
Now we compare our proposed joint SDR turbo receivers with those SDR turbo receivers from \cite{nekuii2011efficient},
which we name as ``Mehran List SDR'' and ``Mehran Single SDR'', respectively. 
The ``Mehran List SDR'' solves SDRs in each iteration while ``Mehran Single SDR''
runs one SDR in the first iteration only. 
For Mehran's methods, we employ same setting as in the paper \cite{nekuii2011efficient}: 
25 randomizations, (at most) 25 preliminary elements in the list, of which 5 elements are used for enrichment. 

\begin{figure}[!htb]
\centering
\centerline{\includegraphics[width=12cm]{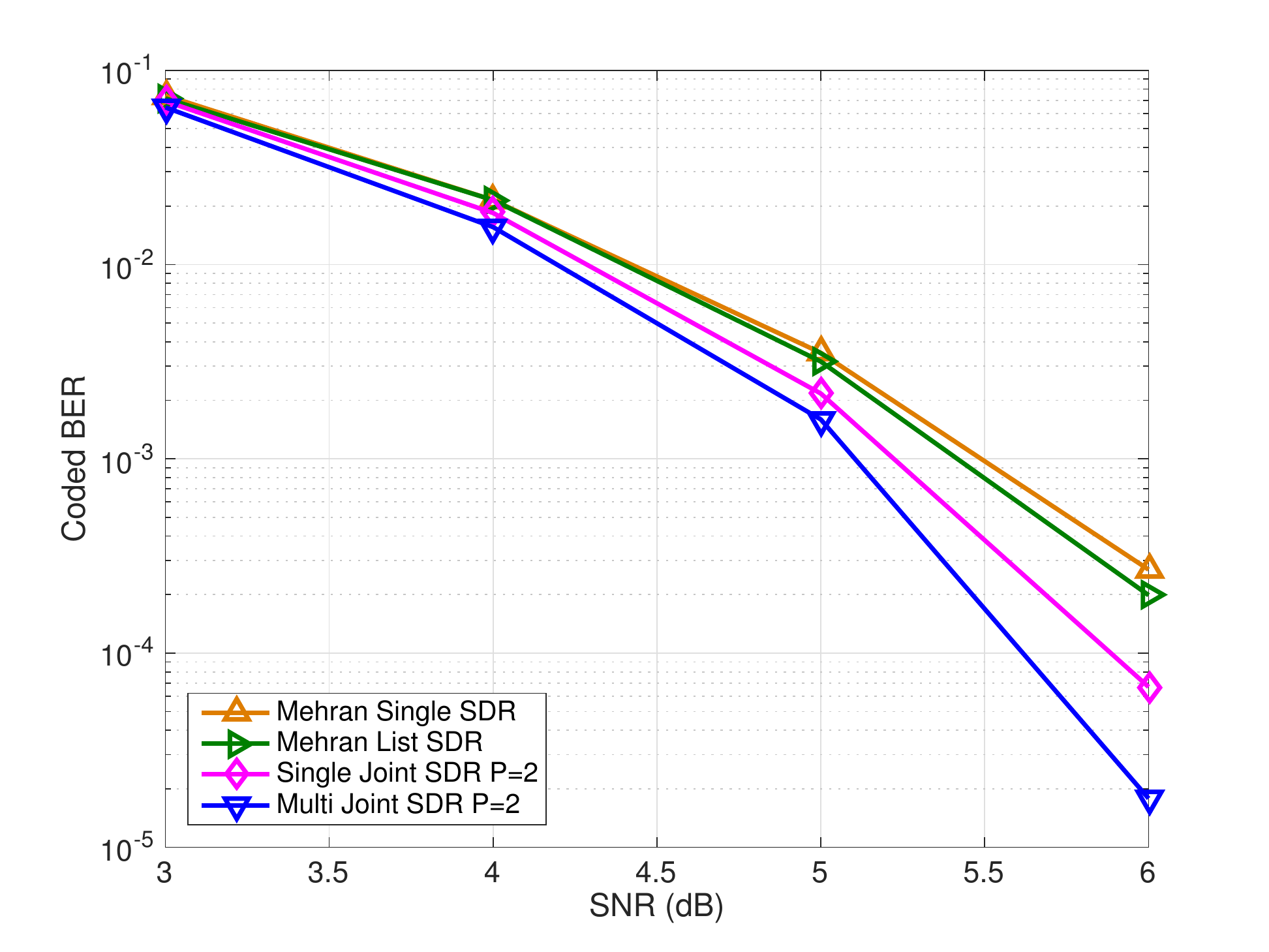}}
\caption{\small{BER comparisons of different Turbo SDR: (256,128) code.}}
\label{fig:ber_comp_all}
\end{figure}

\begin{figure}[!htb]
\centering
\centerline{\includegraphics[width=12cm]{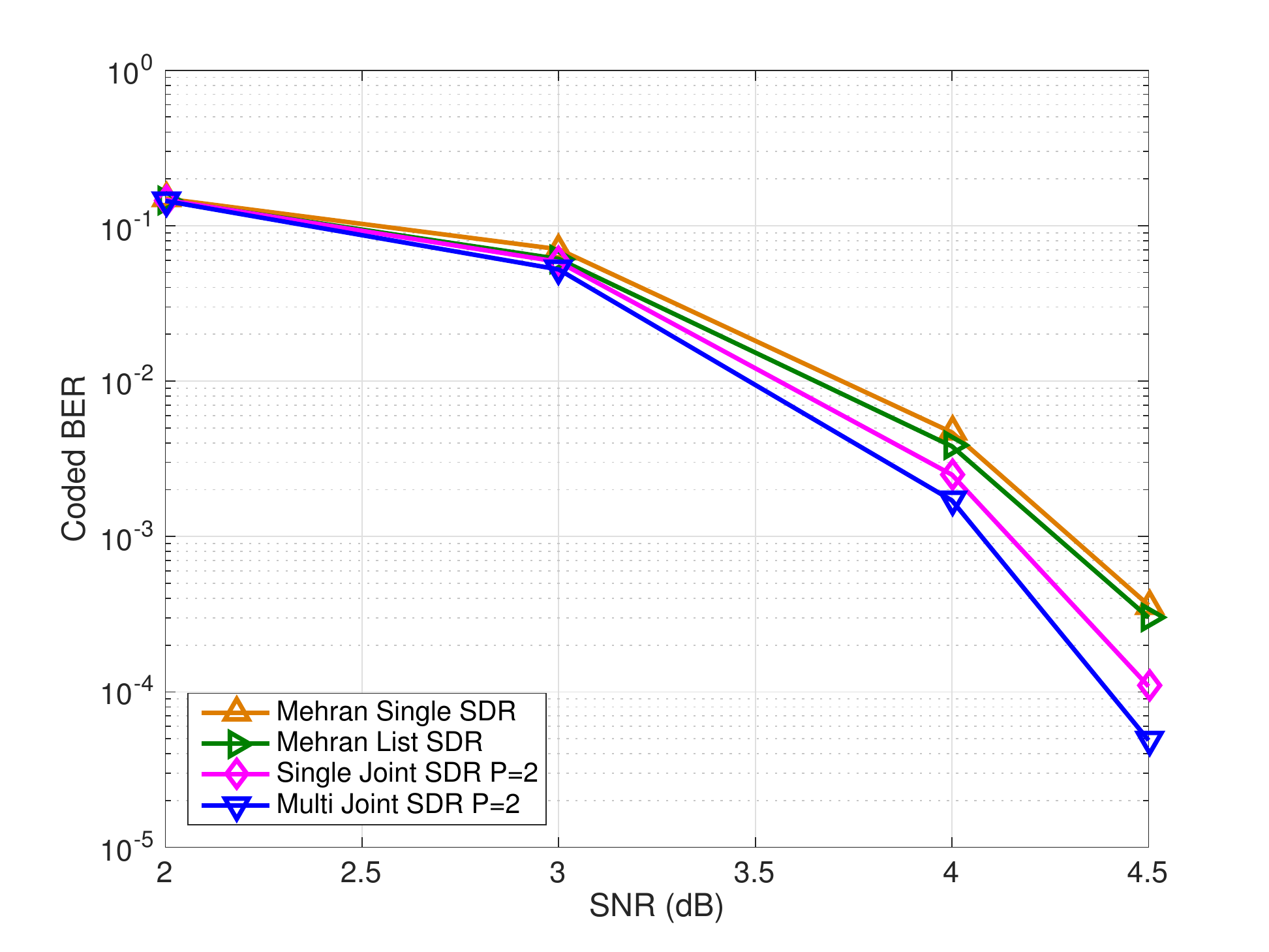}}
\caption{\small{BER comparisons of different Turbo SDR: (1024,512) code.}}
\label{fig:ber_comp_long}
\end{figure}

\begin{figure}[!htb]
\centering
\centerline{\includegraphics[width=12cm]{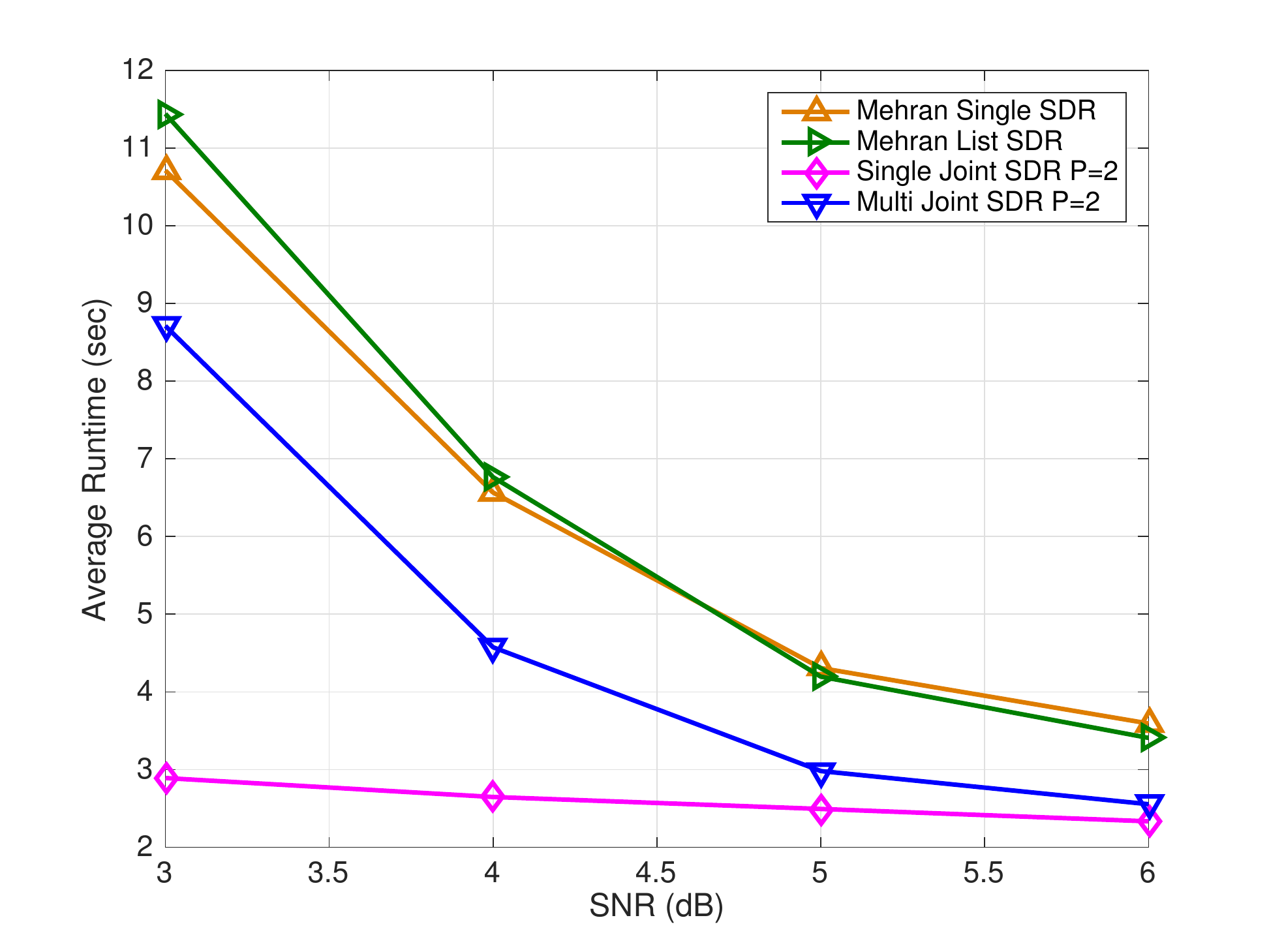}}
\caption{\small{Average runtime comparisons of different Turbo SDR: (256,128) code.}}
\label{fig:time_comp_all}
\end{figure}

We use a (256,128) code in Fig.~\ref{fig:ber_comp_all} while a longer (1024,512) code in Fig.~\ref{fig:ber_comp_long}.
All BER curves are plotted after the 3rd iteration of turbo processing. 
For our joint SDR turbo receivers, Hamming radius $P=2$ for list generation. 
The performance advantage of our receivers is clear around BER = 1e-4. 
The average runtime of different turbo SDR receivers are plotted in Fig.~\ref{fig:time_comp_all}
by using SDPT3 solver in CVX \cite{cvx}. 
We note that this runtime represents the simulation time spent for each codeword, not each MIMO block. 
Thus, it involves solving multiple SDRs for Mehran's methods. 
Because we use early termination, the runtime becomes less and less at higher SNR regime. 
It is clear that our proposed receivers incurred lower complexity than Mehran's receivers, 
especially the single joint SDR turbo receiver that consumed much less time compared to other receivers in low SNR regime.

\section{Conclusion} \label{sec:con}
This work introduces joint SDR detectors integrated with code constraints for MIMO systems. 
The joint ML-SDR detector takes advantage of FEC code information in the detection stage,
and it demonstrates significant performance gain compared to the SDR receiver without code constraints.
We then modify the joint ML-SDR to make it joint MAP-SDR for iterative turbo receiver. 
Our joint MAP-SDR turbo receiver performs equally well as the full list turbo receiver, but at a lower time complexity. 
The proposed turbo receiver is further simplified so that only one SDR is solved in the first iteration per codeword,
and the list generation in subsequent iterations is based on a combination of initial LLRs and decoder feedback LLRs. 
This simplified scheme incurs slight performance degradation, but the complexity is greatly reduced. 
Last but not the least, we remark that the concept of joint receiver design \cite{wang2017galois} can be very effective 
when there exist RF imperfections, such as phase noise \cite{wang2017phase}. 

\bibliographystyle{IEEEtran}
\bibliography{IEEEabrv,mySDRbibfile}
%
%
%

\end{document}